**Percentile Ranks and the Integrated Impact Indicator (I3)**

Loet Leydesdorff [1] and Lutz Bornmann [2]

In a brief communication to this journal, entitled "Basic Properties of Both Percentile Rank Scores and the I3 Indicator," Rousseau (in press) proposed to define percentile classes in the case of the Integrated Impact Indicator (I3) right-open so that the largest number in a set always belongs to the highest percentile. He argued that "this completely solves the issue raised by Leydesdorff & Bornmann (2011) that in the case of small numbers (e.g., reviews), papers would for arithmetic reasons have lower percentile values."

In Leydesdorff & Bornmann (2011, at p. 2137), we proposed to add 0.9 to the count (i.e., count + 0.9) because otherwise one can expect undesirable effects for datasets smaller than 100. For example, if a journal with many articles publishes only 10 reviews each year, the highest possible percentile of reviews within this set would be the 90th (i.e., 9 of 10) whereas this could be the 99th (i.e., 9.9 of 10) when 0.9 is added to the count. Elaborating on this correction—which has marginal effects for numbers in the set larger than 100—Rousseau proposed more radically to include the citation frequency of the paper under study in the count so that in this example the highest possible rank for a review would always be 100%. In formula format, the counting rule that the percentile is determined by the number of items with lower citation rates than the item under study ($x_i < x$ ; $i = 1,\ldots, n$) is replaced with $x_i \leq x$ ; $i = 1,\ldots, n$.

[1] University of Amsterdam, Amsterdam School of Communication Research (ASCoR), Kloveniersburgwal 48, 1012 CX Amsterdam, The Netherlands; loet@leydesdorff.net.
[2] Max Planck Society, Administrative Headquarters, Hofgartenstrasse 8, D-80539 Munich, Germany; bornmann@gv.mpg.de.

Since the counting rule employed for computing percentile values is not uniquely determined (Hyndman & Fan, 1996; Sheskin, 2011, at pp. 40, 120-122),[3] we accepted Rousseau's suggestion as a further improvement, implemented it into the program for computing *I3* in Web-of-Science data (at http://www.leydesdorff.net/software/i3), and began to use it in a recent study (Bornmann & Leydesdorff, in press). However, Zhou *et al*. (in preparation) noted inconsistencies in the empirical application of Rousseau's revision. For example, if one would have a set of nine uncited papers and one with citation, the uncited papers would all be placed in the 90$^{th}$ percentile rank. A lowly-cited document set would thus be advantaged when compared with a highly-cited one.

Rousseau (*personal communication*, Dec. 23, 2011) suggested disregarding the zero-counts in this case. We followed this suggestion, placed all non-cited items in the zero$^{th}$ percentile rank, and re-analyzed the set of 65 LIS journals (JCR 2009) studied by Leydesdorff & Bornmann (2011) in considerable detail. In Table 1, we use the values provided in their Table 4 (at p. 2139) for the comparison: 15 journals of LIS are listed with highest values on *I3* (expressed as percentages of the sum of *I3*) compared with IFs (2009), total citation, and %*I3* on the bases of the six percentile rank classes used by the NSF (National Science Board, 2010; cf. Bornmann & Mutz, 2011).

---

[3] In most cases any differences obtained between the various methods which be employed to identify a score at a specific percentile will be of little or no practical consequence (Sheskin, 2011, at p. 120).

**Table 1**: Rankings of 15 LIS journals with highest values on *I3* (expressed as percentages of the sum) compared with *IF*s, total citations, and % *I3*(*6PR*) with different calculation rules for the percentiles.

| Journal | N of papers (a) | Total citations (b) | % I3 (L&B)* (c) | %I3 (Rousseau)** (d) | % I3 (quantiles) (e) | % I3(6PR) (L&B)* (f) | %I3(6PR) Rousseau** (g) | % I3(6PR) (quantiles) (h) | IF 2009 (i) |
|---|---|---|---|---|---|---|---|---|---|
| *J Am Soc Inf Sci Technol* | 375 | 1975 [1] | 9.72 [1] | 7.32 [2] | 9.73 [1] | 8.63 [1] | 8.45 [1] | 8.64 [1] | 2.300 [7] |
| *Scientometrics* | 258 | 1336 [3] | 7.23 [2] | 5.20 [4] | 7.24 [2] | 6.37 [2] | 6.19 [2] | 6.35 [2] | 2.167 [10] |
| *J Amer Med Inform Assoc* | 199 | 1784 [2] | 6.80 [3] | 4.53 [5] | 6.80 [3] | 6.15 [3] | 5.79 [3] | 6.11 [3] | 3.974 [2] |
| *Inform Process Manage* | 221 | 921 [4] | 6.14 [4] | 4.41 [6] | 6.14 [4] | 4.90 [4] | 4.94 [4] | 4.92 [4] | 1.783 [15] |
| *Inform Management* | 117 | 822 [6] | 4.01 [5] | 2.63 [7[ | 4.01 [5] | 3.35 [5] | 3.22 [7] | 3.33 [5] | 2.282 [8] |
| *Int J Geogr Inf Sci* | 120 | 446 [9] | 3.14 [6] | 2.32 [8] | 3.14 [6] | 2.55 [6] | 2.62 [8] | 2.52 [6] | 1.533 [17] |
| *MIS Quart* | 66 | 847 [5] | 2.61 [7] | 1.61 [21] | 2.61 [7] | 2.34 [7] | 2.17 [11] | 2.32 [7] | 4.485 [1] |
| *J Manage Inform Syst* | 82 | 496 [8] | 2.60 [8] | 1.76 [15] | 2.60 [8] | 2.31 [8] | 2.20 [10] | 2.28 [8] | 2.098 [11] |
| *J Health Commun* | 90 | 380 [10] | 2.52 [9] | 1.80 [14] | 2.51 [9] | 2.04[10a] | 2.02 [14] | 2.04 [10a] | 1.344 [22] |
| *J Acad Libr* | 127 | 252 [19] | 2.50 [10] | 2.15 [9] | 2.51 [10] | 2.05 [9] | 2.24 [9] | 2.06 [9] | 1.000 [26] |
| *J Inform Sci* | 102 | 355 [13] | 2.43 [11] | 1.88 [12] | 2.43 [11] | 1.98 [11] | 2.04 [13] | 1.99 [12] | 1.706 [16] |
| *J Comput-Mediat Commun* | 108 | 374 [11] | 2.37 [12] | 1.89 [11] | 2.37 [12] | 2.04 [10b] | 2.09 [12] | 2.04 [10b] | 3.639 [3] |
| *J Informetr* | 66 | 598 [7] | 2.28 [13] | 1.49 [24] | 2.28 [13] | 2.04 [10c] | 1.97 [16] | 2.03 [11] | 3.379 [4] |
| *J Med Libr Assoc* | 114 | 248 [20] | 2.21 [14] | 1.97 [10] | 2.21 [14] | 1.93 [12] | 1.98 [15] | 1.94 [13] | 0.889 [31] |
| *Telecommun Policy* | 96 | 264 [17] | 2.15 [15] | 1.74 [18] | 2.15 [15] | 1.80 [13] | 1.80 [17] | 1.81 [14] | 0.969 [27] |

* Source: Leydesdorff & Bornmann (2011); ** cf. Rousseau (in press).

Table 1 shows the values of % *I3* on the basis of the correction (of + 0.9) proposed by Leydesdorff & Bornmann (2011) in column *c*, by Rousseau (in press) in column *d*, and using the quantiles without a correction in column *e*. Analogously, columns *f* to *h* show these values for using the six percentile ranks (top-1%, 95-99%, 90-95%, 75-90%, 50-75%, and bottom-50%) used in the *Science and Engineering Indicators* of the National Science Board (2010).

The differences between the quantiles and the correction with +0.9 are only in the second decimal of the percentages and negligible ($r = 1.00; p < 0.01; N = 65$). (This gives some confidence that the much smaller differences generated by using different calculation rules for quantiles will have no significant effect on rankings using *I3*.) However, the differences with the values provided in column *d* based on the normalization suggested by Rousseau (in press) are considerable although the correlation coefficient is still high ($r = 0.70; p < 0.01; N = 65$). For example, *JASIST* would lose its first position in this ranking to *The Scientist*, and is now on the second position.

*The Scientist* contained 392 citable items in 2008 and 2009, of which 352 (98.1%) were not cited at the time of our download (February 2011). Using Rousseau's counting rule with the improvement specified above, the journal obtains a %*I3* of 7.50, which is above the 7.32% attributed to *JASIST* (in columne *d*). Using quantiles, however, *The Scientist* is rated 33$^{rd}$ with a %*I3* of 1.00, and therefore not listed among the top-15 journals in Table 1.[4] Figure 1 shows the distributions when the 40 remaining documents of *The Scientist* (which were cited at least once) are rated in the six percentile rank classes comparatively. (The 352 papers with no citations would be placed in the lowest category if they were counted in.)

---

[4] In the case of six percentile ranks, %I3(6PR) in column *g* is 3.90%. *The Scientist* would then be ranked 5$^{th}$.

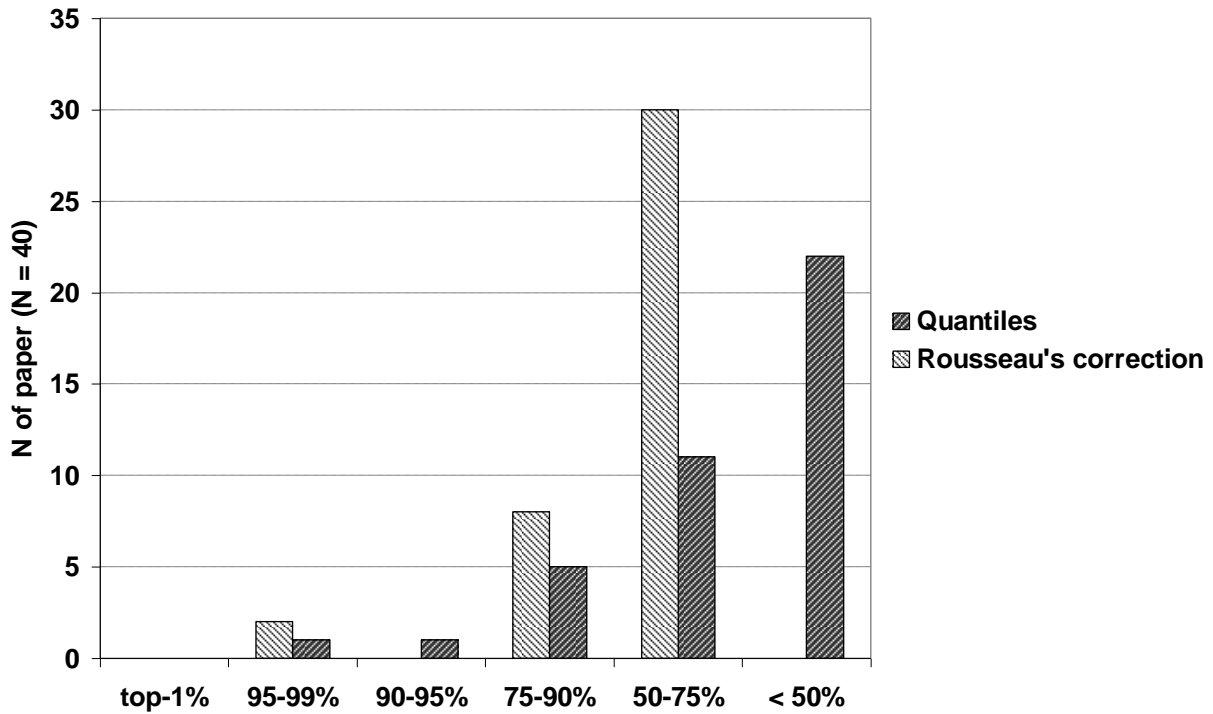

**Figure 1**: The distribution among six percentile rank classes (Bornmann & Mutz, 2011; National Science Board, 2010) for 40 citable documents in *The Scientist* which were cited at least once.

In summary, we regret with hindsight our suggestion to deviate from quantiles (however computed) as a basis for the ranking because Rousseau's contribution makes clear that we may have opened a box of Pandora allowing for generating a parameter space of other possibilities. The mathematical discussion about other possibilities easily obscures our central message that one is not allowed nor does one have to use central tendency statistics for analyzing citation distributions (Seglen, 1992). Nonparametric statistics is available for the measurement and the testing of the statistical significance of differences. Notwithstanding our reservations, we extended the program at http://www.leydesdorff.net/software/i3 with the three options available; the quantiles without a correction are now the default option.

*I3* is an *impact* indicator which can be used as an alternative to parametric statistics such as the ratio of citations over publications (*c/p*) or the *IF* (Rousseau & Leydesdorff, 2011). The advantage is that one accounts for the expected skewness of citation distributions using non-parametric statistics. More recently, both the *SCImago Institutions Rankings* (at http://www.scimagoir.com/pdf/sir_2011_world_report.pdf) and the *Leiden Ranking 2011/2012* (at http://www.leidenranking.com/ranking.aspx) used the top-10% most-highly cited papers as an *excellence* indicator (Bornmann & Leydesdorff, 2011).

In the case of these excellence indicators, only two percentile rank classes are distinguished for the evaluation (Rousseau, 2011 and in press). Both excellence and impact indicators can be tested against expectation or in terms of differences between two ranks using the *z*-test for independent proportions (Bornmann *et al.*, 2011; Leydesdorff & Bornmann, in preparation). In short, Rousseau's stated preference to solve problems first mathematically (as in an "ideal gas") provided us with an empirically testable hypothesis.

**Acknowledgements**
We are grateful to Ronald Rousseau and Ping Zhou for comments and suggestions.